\documentclass[11pt,twoside]{article}
\usepackage{asp2010}

\resetcounters

\begin{document}

\title{Rotational properties of single and wide binary subdwarf B stars}
\author{S. Geier$^1$, U. Heber$^1$, H. Edelmann$^1$, R. Napiwotzki$^2$, L. Morales-Rueda$^3$
\affil{$^1$Dr. Karl Remeis-Observatory \& ECAP, Astronomical Institute,
Friedrich-Alexander University Erlangen-Nuremberg, Sternwartstr. 7, D 96049 Bamberg, Germany}
\affil{$^2$Centre of Astrophysics Research, University of Hertfordshire, College Lane, Hatfield AL10 9AB, UK}
\affil{$^3$Department of Astrophysics, Faculty of Science, Radboud University Nijmegen, P.O. Box 9010, 6500 GL Nijmegen, NE}}

\begin{abstract}
We measured projected rotational velocities of more than a hundred apparently single sdBs. A comparison is made with sdB stars in binary systems with orbits so wide, that tidal interaction becomes negligible. All of these stars are slow rotators ($v_{\rm rot}\sin{i}<10\,{\rm km\,s^{-1}}$) with EC\,22081$-$1916 being the only exception. This single star has the highest projected rotational velocity ever measured for an sdB ($v_{\rm rot}\sin{i}=163\,{\rm km\,s^{-1}}$) and might have been formed by a merger event. The merger of a red-giant core and a low-mass, main-sequence star or substellar object during a common envelope phase fits particularly well with observations. The implications of our results for hot subdwarf formation are briefly discussed.
\end{abstract}

\section{Introduction}

The formation of hot subdwarf stars (sdBs) is still unclear. Several different single star and close binary formation scenarios are discussed to explain the enhanced mass loss in the red-giant phase necessary to form sdBs. As an alternative, the merger of compact objects (e.g. helium white dwarfs, He-WDs) has been proposed as a possible origin for single sdBs (for the most recent review see Han et al. these proceedings). Whether and how the progenitor star loses or gains angular momentum should affect the rotational properties of the resulting sdB. Determining the $v_{\rm rot}\sin{i}$-distribution of sdBs should therefore be useful to shed light on their formation. 

While the rotational properties of horizontal branch stars both in 
globular clusters and in the field are thoroughly examined \citep[e.g.][]{behr03}, there is no systematic study for extreme horizontal branch (EHB) stars yet. Most of the sdB stars where $v_{\rm rot}\sin{i}$-measurements are available, are slow rotators \citep[$<10\,{\rm km\,s^{-1}}$, e.g.][]{heber00, edelmann01}. Here we determine the projected rotational velocities of more than a hundred sdB stars by measuring the broadening of metal lines. This study complements the investigation of the rotational properties of close binary sdBs \citep{geier10}.

\begin{figure}
\begin{center}
  \includegraphics[width=8cm]{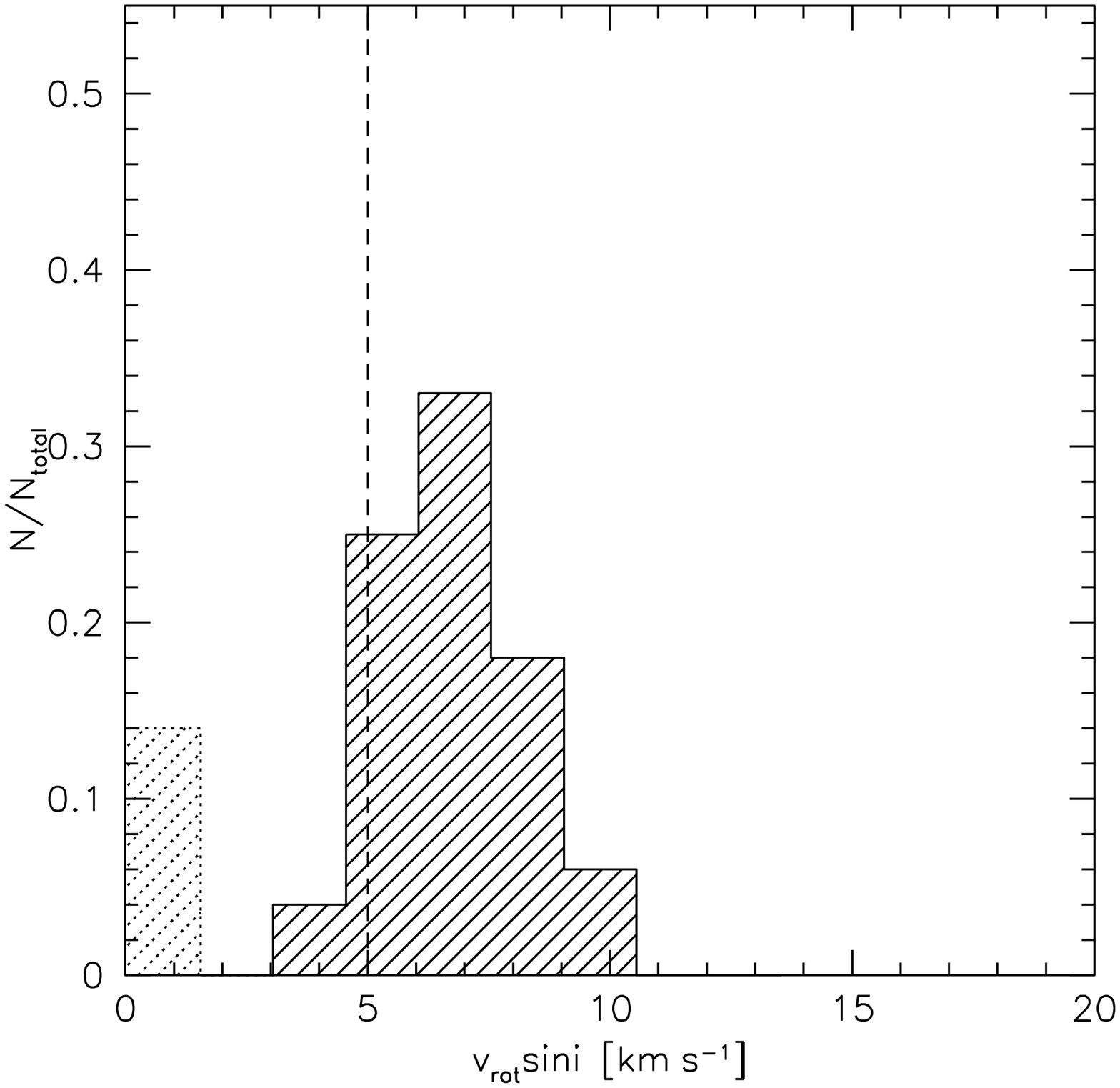}
  \includegraphics[width=8cm]{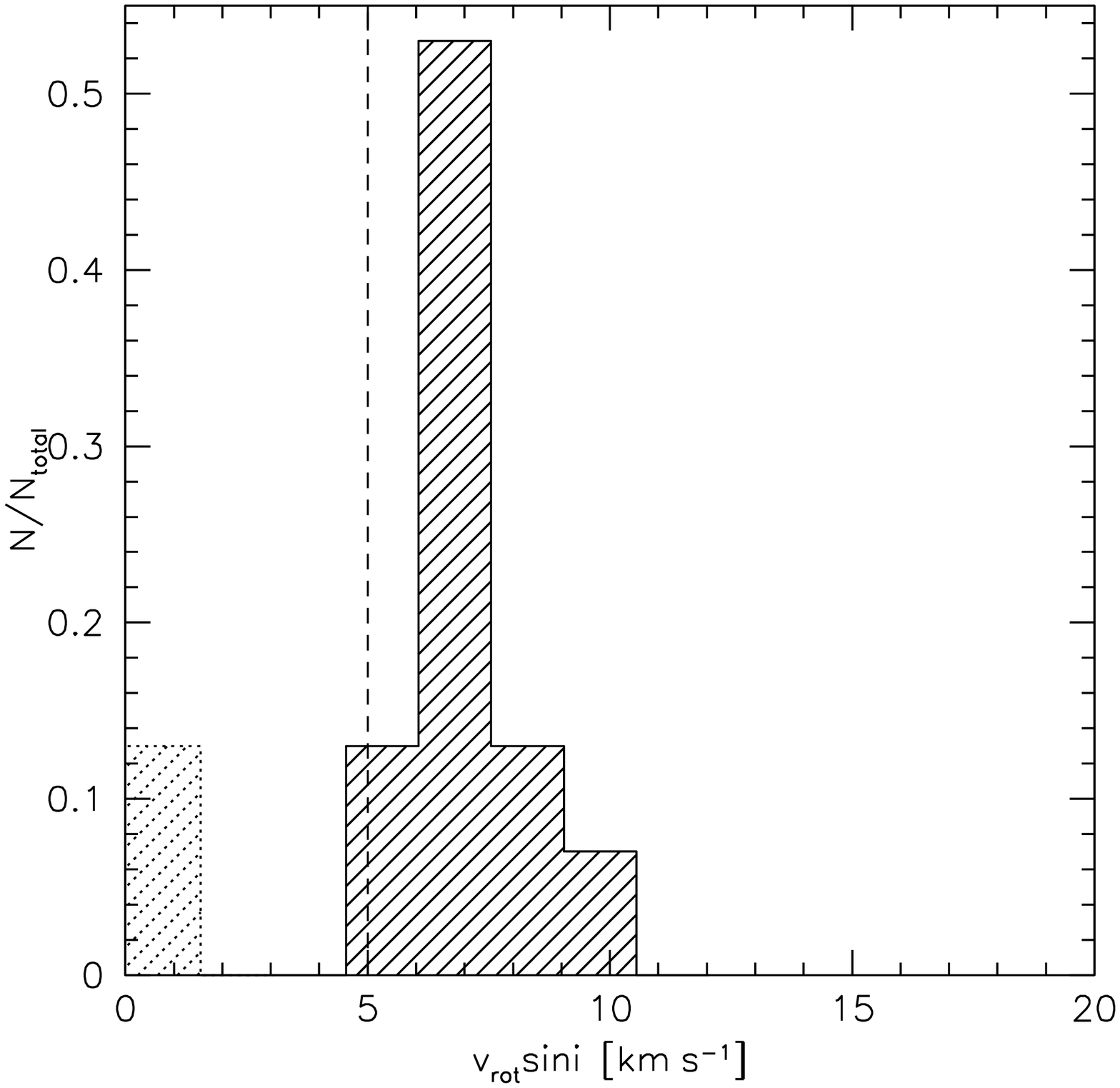}
\end{center}
\caption{Upper panel: Distribution of ${v_{\rm rot}\sin\,i}$ for 72 single stars from our sample. Lower panel: Distribution of ${v_{\rm rot}\sin\,i}$ for 15 sdBs with visible companions.}
\label{fig:rotsingle}
\end{figure}

\section{Data Analysis}

We derived metal abundances and projected rotational velocities of sdBs using high re\-solution spectra from the SPY sample \citep{lisker05} and a sample of bright sdBs (see Geier et al. these proceedings). For a standard set of up to 187 unblended metal lines from 24 different ions in the wavelength range from $3700$ to $6000\,{\rm \AA}$ a model grid with appropriate atmospheric parameters and different elemental abundances was generated using LTE atmospheric models \citep{heber00} and the spectrum synthesis code LINFOR \citep{lemke97}. A simultaneous fit of elemental abundance, projected rotational velocity and radial velocity was then performed separately for each line using the FITSB2 routine \citep{napiwotzki04}.

\section{Projected Rotational Velocities}

We selected 103 sdBs, which do not show RV variations in multi-epoch, high resolution spectroscopy. In most cases no spectral signature of a companion was found. However, 15 sdBs show features indicative a low-mass main-sequence companion (e.g. Mg\,{\sc i}). Such systems might have been formed by stable Roche lobe overflow (see \O stensen et al. these proceedings). Since such a close binary interaction may influence the rotation of the sdB, these stars are treated separately. 

We ended up with 72 single sdBs, where the ${v_{\rm rot}\sin\,i}$ could be reasonably well constrained. Fig.~\ref{fig:rotsingle} (upper panel) shows the distribution of ${v_{\rm rot}\sin\,i}$ binned to the average measurement error ($1.5\,{\rm km\,s^{-1}}$). The distribution is very uniform and shows a prominent peak at $6-8\,{\rm km\,s^{-1}}$. Because we measure the projected rotation, the true rotational velocities of most stars in the sample may be higher ($7-8\,{\rm km\,s^{-1}}$). The distribution of sdBs with visible companions (Fig.~\ref{fig:rotsingle}, lower panel), even when taking into account the much smaller sample size, is very similar. We therefore conclude that the rotational properties of those sdBs are not significantly affected by the presence of main sequence companions.

\begin{figure}
\begin{center}
  \includegraphics[width=8cm]{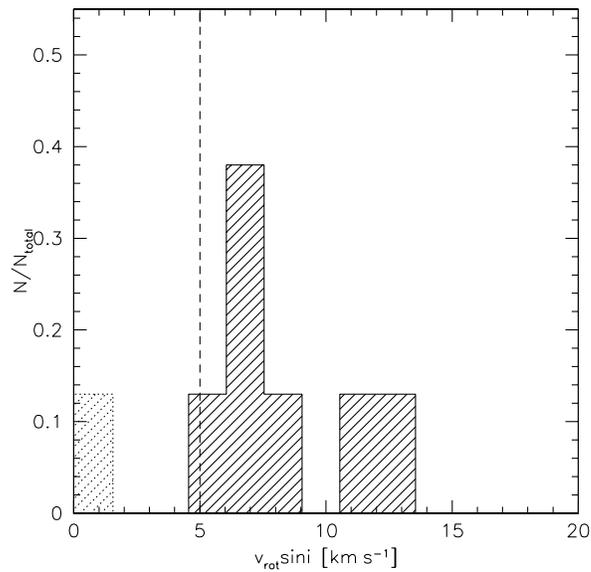}
\end{center}
\caption{Distribution of ${v_{\rm rot}\sin\,i}$ for 8 radial velocity variable sdBs with long orbital periods.}
\label{fig:rotRV}
\end{figure}

\begin{figure}
\begin{center}
  \includegraphics[width=10cm]{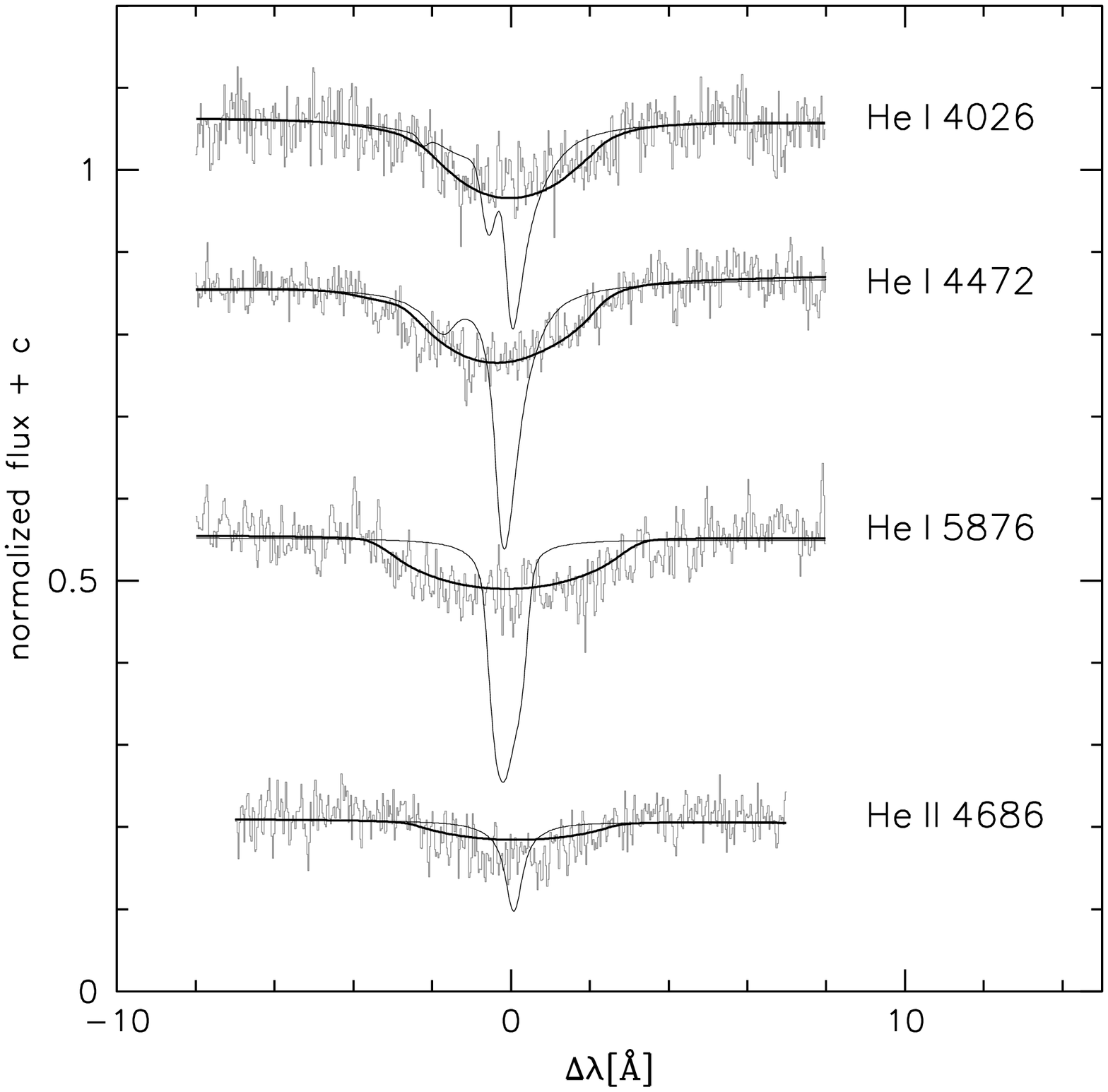}
  \includegraphics[width=10cm]{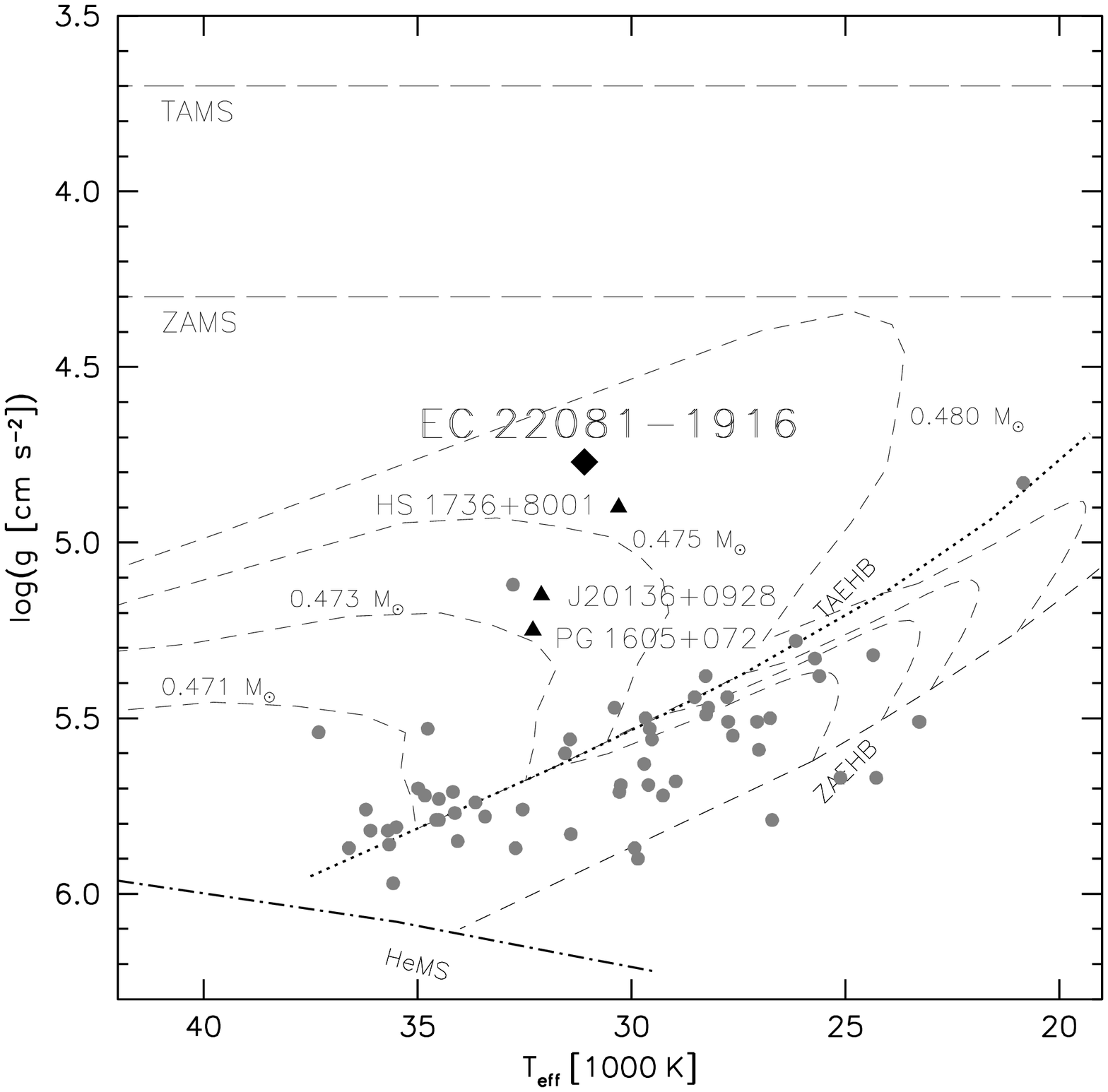}
\end{center}
\caption{Upper panel: Fit of models to helium lines of EC\,22081$-$1916. The thin solid line marks models without rotational broadening, the thick solid line the best fitting model spectrum. The extreme rotational broadening of the lines is obvious. Lower panel: $T_{\rm eff}-\log{g}$ diagram. The grey circles mark sdBs from the SPY project. The low gravity sdBs HS\,1736+8001 \citep{edelmann03}, PG\,1605+072 \citep{heber99} and J20136+0928 \citep{oestensen11} are plotted as triangles. EHB evolutionary tracks are taken from \citet{dorman93}. The location of the main sequence is indicated by the long-dashed horizontal lines \citep{geier11}.}
\label{fig:rotHe}
\end{figure}
 
In \citet{geier10} we showed that the ${v_{\rm rot}\sin\,i}$ distribution of sdBs in close binary systems is strongly modified by the tidal interaction with their companions, but that this influence becomes negligible if the orbital periods of the binaries become longer than $\simeq1.2\,{\rm d}$. We selected all seven binaries with periods longer than $1.2\,{\rm d}$, where tidal synchronisation is not established and the long-period system LB\,1516, for which orbital parameters are not yet known. The lower limit for the period is of the order of days \citep{edelmann05}. Fig.~\ref{fig:rotRV} shows the associated distribution. Given the small sample size and the fact that two stars have somewhat higher ${v_{\rm rot}\sin\,i}=10-12\,{\rm km\,s^{-1}}$ the distribution is again very similar to the distributions shown in Fig~\ref{fig:rotsingle}. SdBs in close binaries obviously rotate in a similar way as single stars or sdBs with visible companions given that the orbital period is sufficiently long.

\section{EC\,22081$-$1916 - A merger remnant?}

EC\,22081$-$1916 is an outlier in our sample because it displays strongly broadened line profiles. It was discovered in the course of the Edinburgh-Cape blue object survey and classified as an sdB star by \citet{copperwheat11}. The projected rotational velocity was derived from the Balmer and helium lines. In order to fit these lines, a very high rotational broadening of $v_{\rm rot}\sin{i}=163\pm3\,{\rm km\,s^{-1}}$ is necessary (see Fig.~\ref{fig:rotHe}, upper panel). The resulting effective temperature $T_{\rm eff}=31\,100\pm1000\,{\rm K}$, and helium abundance $\log{y}=-1.97\pm0.02$ are typical for sdB stars, whereas the surface gravity of $\log{g}=4.77\pm0.10$ is unusually low for the effective temperature in question (see Fig.~\ref{fig:rotHe}, lower panel).

No significant RV variations were measured. After excluding all possible alternative scenarios (MS star, pulsator, close binary) we conclude that EC\,22081$-$1916 is the first single sdB star which is rapidly rotating \citep[for details see ][]{geier11}. EC\,22081$-$1916 thus might have been formed by a merger event. He-WD mergers are believed to have very small envelope masses and are expected to be situated at the very blue end of the EHB. Both are at variance with the position of EC\,22081$-$1916 in the ($T_{\rm eff}-\log{g}$)-diagram. \citet{justham10} proposed that the merger of a close binary system consisting of an sdB and a He-WD may form a single helium enriched sdO. EC\,22081$-$1916, however, is helium-deficient.

The merger of a red-giant core and a low-mass, main-sequence star or substellar object during a common envelope phase may also lead to the formation of a rapidly rotating hot subdwarf star \citep{politano08}. This scenario fits particularly well with observations, because it provides a natural explanation for the thick hydrogen envelope. EC\,22081$-$1916 is unique among $\simeq100$ slowly rotating sdB stars analysed so far.

\section{Discussion}

The main conclusion to be drawn from our results is that the rotational properties of single sdBs, and both wide and close binary sdBs with orbital periods of the order of a few days are very similar. All of these objects have very small $v_{\rm rot}\sin{i}$. However, according to the standard binary formation scenario for sdB stars, these stars should have been formed in quite different ways. In binary systems, mass transfer is proposed to be responsible for the loss of the hydrogen envelope in the red-giant phase. Depending on the initial conditions, this mass loss can be either stable or unstable. Since envelope mass and angular momentum are lost by the sdB progenitor in both cases, it is not surprising that the remnant rotates slowly.

Single sdB stars on the other hand have been proposed to be the remnants of the merger of two He-WDs. Although the angular momentum evolution of the merger product is unknown in detail, it is surprising to find single sdBs stars rotating with basically the same $v_{\rm rot}\sin{i}$ as in binaries despite the totally different formation channel. Whether this intriguing behaviour can be explained in the framework of the merger channel remains to be seen (see Podsiadlowski et al. these proceedings). 

EC\,22081$-$1916 is unique among $\simeq100$ slowly rotating sdB stars and may be the rare outcome of a common envelope merger event.


\end{document}